\documentclass[CEJP,PDF]{cej} 
\usepackage{layout}
\usepackage{amsmath}
\usepackage{textcomp}
\usepackage{graphicx}

\title{Photonic realization of the relativistic Kronig-Penney model and relativistic Tamm surface states}

\articletype{Research Article}

\author{Stefano Longhi\inst{1}\email{longhi@fisi.polimi.it}
       }

\institute{
     \inst{1} Dipartimento di Fisica, Politecnico di Milano\\
     Piazza Leonardo da Vinci 32, I-20133 Milano (Italy)
          }

\abstract{Photonic analogues of the relativistic Kronig-Penney model
and of relativistic surface Tamm states are proposed for light
propagation in fibre Bragg gratings (FBGs) with phase defects. A
periodic sequence of phase slips in the FBG realizes the
relativistic Kronig-Penney model, the band structure of which being
mapped into the spectral response of the FBG. For the semi-infinite
FBG Tamm surface states can appear and can be visualized as narrow
resonance peaks in the transmission spectrum of the grating.}


\begin{document}
\maketitle

\section{Introduction} The Kronig-Penney model for the non-relativistic
Schr\"{o}dinger equation \cite{KP} is a well-known model in
solid-state physics that describes the electronic band structure of
an idealized one-dimensional crystal. The Kronig-Penney model has
served on many occasions as a paradigmatic model to study a wide
variety of physical phenomena, including band structure properties,
localization effects in disordered lattices, electronic properties
of superlattices, Peierls transitions and quark tunnelling in
one-dimensional nuclear models. Relativistic extensions of the
Kronig-Penney model (also referred to as the Dirac-Kronig-Penney
model) have been discussed by several authors (see, for instance,
\cite{RKP,D0,D1,D2,D3,D4,D5,D6,D7,D8,D9,D10} and references
therein), and the impact of relativity on the band structure and
localization, such as shrinkage of the bulk bands with increasing
band number, have been highlighted on many occasions. In earlier
studies, the Dirac-Kronig-Penney model also attracted some attention
and caused a lively debate about the existence of so-called Dirac
surface states, i.e. relativistic surface Tamm states which
disappear in the non-relativistic limit
\cite{D0,D4,SS1,SS2,SS3,SS4,SS5,SS6}. In recent years, there has
been an increased interest in simulations of relativistic quantum
effects using different physical set-ups, and analogues of such
fundamental phenomena as Zitterbewegung  and Klein tunnelling
-rooted in the Dirac equation- have been proposed and demonstrated
for electrons in graphene \cite{graphene1,graphene2} and for matter
waves using trapped cold atoms \cite{coldatoms}. Light propagation
in guiding optical structures has been also shown to provide a
beautiful laboratory system to investigate the classical analogous
of a wide variety of coherent non-relativistic \cite{Longhi09LPR}
and relativistic \cite{R1,R2,R3,R4,R5} quantum phenomena. In
particular, it was recently shown that light propagation in fibre
Bragg gratings (FBGs), i.e. optical fibres with a superimposed
modulation of the refractive index profile, provides an
experimentally accessible laboratory tool to simulate in optics the
massive one-dimensional Dirac equation, and a photonic realization
of the Dirac oscillator (i.e. the relativistic extension of the
quantum harmonic oscillator) has been proposed
\cite{Longhi10OL}.\\
In this work we propose a photonic realization of the relativistic
Kronig-Penney model and relativistic surface Tamm states based on
light propagation in FBGs with phase defects.  Light propagation in
a FBG with a periodic sequence of phase slips is shown to simulate
the relativistic Kronig-Penney model, the relativistic band
structure of which being mapped into the spectral transmission of
the FBG. Similarly, a semi-infinite FBG with phase defects
interfaced with a uniform FBG with a different modulation period is
shown to support Tamm surface states analogous to the relativistic
Tamm states. Such surface states are responsible for narrow
resonance peaks in the transmission spectrum of the grating. The
paper is organized as follows. In Section 2, the photonic
realization of the Dirac-Kronig-Penney model for an
infinitely-extended crystal, based on a superstructure FBG, is
presented. The photonic analogues of relativistic Tamm states in a
semi-infinite lattice model are discussed in Section 3, and  shown
to appear as narrow resonance peaks in the spectral transmission of
the FBG. Finally, the main conclusions are outlined in Section 4.

\section{Photonic realization of the Dirac-Kronig-Penney model}
In this section a photonic realization of the Dirac-Kronig-Penney
model for an infinitely-extended lattice, based on Bragg scattering
of light waves in a superstructure FBG comprising a periodic
sequence of phase slips, is proposed. Let us consider light
propagation in a FBG with a longitudinal effective refractive index
given by
\begin{equation}
 n(z)=n_0+\Delta n \; m(z) \cos \left[ \frac{2 \pi z}{ \Lambda} + \phi(z) \right],
\end{equation}
where $n_0$ is the effective mode index in absence of the grating,
$\Delta n \ll n_0$ is the peak index change of  the grating,
$\Lambda$ is the nominal period of the grating defining the
reference frequency $\omega_B=\pi c/(\Lambda n_0)$ of Bragg
scattering, $c$ is the speed of light in vacuum, and $m(z)$,
$\phi(z)$ are the slow variation, as compared to the scale of
$\Lambda$, of normalized amplitude and phase, respectively, of the
index modulation. The periodic index modulation of the grating leads
to Bragg scattering between two counterpropagating waves at
frequencies close to $\omega_B$. By letting
\begin{equation}
E(z,t)=\varphi_1(z,t) \exp \left[-i \omega_B t +ik_B z+i \phi(z)/2
\right]+ \varphi_2(z,t) \exp \left[ -i \omega_B t -ik_B z-i
\phi(z)/2 \right]+c.c.
\end{equation}
 for the longitudinal electric field amplitude in the fibre, where
$k_B=\pi/\Lambda$, the slowly-varying envelopes $\varphi_1$ and
$\varphi_2$ of counterpropagating waves satisfy the coupled-mode
equations (see, for instance, \cite{Erdogan})
\begin{eqnarray}
i \left[ \frac{\partial}{\partial z} + \frac{1}{v_g} \frac{
\partial}{\partial t}
\right] \varphi_1 & = & \frac{1}{2} \left( \frac{d \phi}{dz} \right) \varphi_1-\kappa(z) \varphi_2 \\
i \left[ -\frac{ \partial}{\partial z} + \frac{1}{v_g} \frac{
\partial}{\partial t} \right] \varphi_2 & = & \frac{1}{2}  \left( \frac{d \phi}{dz} \right) \varphi_2-\kappa(z)
\varphi_1
\end{eqnarray}
where we have set
\begin{equation}
\kappa(z) \equiv \frac{k_B m(z) \Delta n}{2n_0}
\end{equation}
and $v_g \sim c/n_0$ is the group velocity at the Bragg wavelength
$\lambda_B=2 \pi c / \omega_B$. The analogy between Bragg scattering
of counterpropagating light waves in the FBG and the temporal
evolution of two-component spinor Dirac equation in presence of an
electrostatic potential is at best captured by introducing the
dimensionless variables $x=z/Z$ and $\tau=t/T$ , with characteristic
spatial and time scales defined by
\begin{equation}
Z=\frac{2 n_0}{k_B \Delta n} \; , \; \; \; T=\frac{Z}{v_g}.
\end{equation}
After introduction of new envelopes
\begin{equation}
\psi_{1,2}(z)=\frac{\varphi_1(z) \mp \varphi_2(z)}{ \sqrt 2},
\end{equation}
Eqs.(3) and (4) can be cast in the Dirac form
\begin{equation}
i \frac{\partial \psi} {\partial \tau} =-i \sigma_1 \frac{\partial
\psi}{\partial x} \psi +m (x) \sigma_3 \psi +V(x) \psi
\end{equation}
for the spinor wave function $\psi=(\psi_1,\psi_2)^T$, where
\begin{equation}
V(x)=\frac{1}{2}  \frac{d \phi}{dx}
\end{equation}
 and $\sigma_{1,3}$ are the Pauli matrices,
defined by
\begin{equation}
\sigma_1= \left(
\begin{array}{cc}
0 & 1 \\
1 & 0
\end{array}
\right) \; , \; \; \sigma_3= \left(
\begin{array}{cc}
1 & 0 \\
0 & -1
\end{array}
\right).
\end{equation}
In its present form, Eq.(8) is analogous to the one-dimensional
Dirac equation, written in natural units $\hbar=c=1$, in the
presence of an external electrostatic potential $V(x)$,  $m(x)$
playing the role of a dimensionless (and generally space-dependent)
rest mass.\\
The Dirac-Kronig-Penney model for an infinitely-extended lattice
corresponds to a constant mass $m(x)=m_0$ and to a potential $V(x)$
given by the superposition of equally-spaced $\delta$-like barriers,
namely \cite{RKP}
\begin{equation}
V(x)=V_0 \sum_{n=-\infty}^{\infty} \delta(x-na)
\end{equation}
where $V_0>0$ is the area of the barrier and $a$ is the lattice
period. Stationary solutions $\psi(x,\tau)=\psi_0(x) \exp(-iE \tau)$
to the Dirac equation (8) in the periodic potential (11) with energy
$E$ are of Bloch-Floquet type, i.e. $\psi_0(x+a)=\psi_0(x)
\exp(iqa)$, where $q$ is the Bloch wave number which varies in the
first Brilloiun zone ($-\pi/a \leq q < \pi/a$). The corresponding
energy spectrum is composed by a set of allowed energy bands
$E=E(q)$, which are defined by the following implicit equation (see,
for instance, \cite{SS4})
\begin{equation}
\cos(qa)=\cos(V_0) \cos( \kappa a)+\frac{E}{\kappa} \sin(V_0)
\sin(\kappa a),
\end{equation}
where we have set
\begin{equation}
\kappa=\sqrt{E^2-m_0^2}.
\end{equation}
Equation (12) defines the dispersion relation of the relativistic
Kronig-Penney model, which has been investigated by several authors
(see, for instance, \cite{RKP,D1,D4}). The ordinary non-relativistic
limit of the Kronig-Penney model is attained from Eqs.(12) and (13)
for $V_0 \ll 1$ and for energies $E$ close the $m_0$, for which the
energy-momentum relation (13) reduces to the non-relativistic one
[$E \simeq m_0+\kappa^2/(2m_0)$]; in this regime, the dispersion
relation (12) reduces to $\cos(qa)=\cos(\kappa a)+(m_0 V_0/\kappa)
\sin(\kappa a)$, which is the ordinary dispersion relation
encountered in the non-relativistic Kronig-Penney model. For larger
energies $E$ but still for a low barrier area $V_0 \ll 1$,
non-relativistic effects come into play as perturbative effects,
which modify positions and widths of the allowed energy bands.
Non-relativistic effects deeply modify the band structure of the
crystal for potential strengths $V_0$ of the order $\sim 1$. In
particular, if $V_0$ is an integer multiple of $\pi$, all band gaps
disappear and the dispersion relation reduces to the one of a
relativistic free particle [Eq.(8) with $V(x)=0$],
as if the $\delta$-barriers were absent.\\
In our photonic context, owing to Eq.(9) the Dirac-Kronig-Penney
model simply corresponds to an infinitely-extended uniform FBG with
a superimposed periodic sequence of lumped phase slips of equal
amplitude $\Delta \phi=2 V_0$ and spaced by the distance $a$. The
circumstance that the effects of the $\delta$ barriers disappear in
the Dirac-Kronig-Penney model when $V_0$ is an integer multiple
$\pi$ is simply due to the fact that, under such a condition, the
phase slips are integer multiplies of $2 \pi$, and thus the grating
has no phase defects and mimics the dynamics of a one-dimensional
free relativistic Dirac particle. It should be noticed that, in
another physical context, superstuctured FBGs comprising a periodic
sequence of $\pi$ phase slips, corresponding to the special case
$V_0=\pi/2$, have been recently proposed and demonstrated to realize
light slowing down \cite{Longhi1,Longhi2}; however, their connection
to the Dirac-Kronig-Penney model was not noticed. It should be also
noticed that any FBG structure has a finite length $L$, and thus
some kind of lattice truncation should be introduced in the
idealized model. As shown e.g. in \cite{Longhi1,Longhi2},
apodization of the FBG amplitude profile $m(x)$, obtained by slowly
decreasing $m(x)$ from its constant value $m(x)=m_0$ in the uniform
grating region to zero at $x \rightarrow \pm \infty$, enables to
adiabatically inject and eject light waves in the uniform grating
region avoiding truncation effects. Correspondingly, the band
structure of the Dirac-Kronig-Penney model is mapped into the
alternation of stop/transmission bands observed in
spectrally-resolved transmission measurements of the FBG. The
spectral transmission ($t$) and reflection ($r$) coefficients for
the Dirac equation with $m(x) \rightarrow 0$ at $x \rightarrow \pm
\infty$, corresponding to left wave incidence, are defined from the
scattering solution to (8) with the following asymptotic behavior
\begin{eqnarray}
\psi(x, \tau) \sim \left\{
\begin{array}{cc}
\left(
\begin{array}{c}
1 \\ 1
\end{array}
\right) \exp[i E x -iE \tau -i \phi(x)/2]+r(E) \left(
\begin{array}{c}
-1 \\ 1
\end{array}
\right)
\exp[-iE x -iE \tau+i\phi(x)/2] & x \rightarrow -\infty \\
t(E) \left(
\begin{array}{c}
1 \\ 1
\end{array}
\right) \exp[i E x -iE \tau-i \phi(x)/2] & x \rightarrow \infty
\end{array}
\right.
\end{eqnarray}
Note that, in our photonic analogue, the energy $E$ of the scattered
waves corresponds to the (normalized) frequency detuning of light
waves, propagating in the grating, from the Bragg reference
frequency $\omega_B$.\\
As an example, Fig.1(c) shows the numerically-computed spectral
power transmission $|t(E)|^2$ (in dB units) for a uniform FBG
comprising a sequence of $\pi$ phase slips [see Fig.1(b)] and with a
super-Gaussian apodizion profile $m(x)$ shown in Fig.1(a); parameter
values are $V_0=\pi/2$, $a=2$, $m_0=1$ and $L=50$. The power
spectral transmission has been computed using a standard transfer
matrix method (see, for instance, \cite{Erdogan}). The shaded areas
in Fig.1(c) indicate the stop bands of the ideal (infinitely-long)
Dirac-Kronig-Penney lattice as predicted by Eq.(12). In physical
units, for typical parameter values $n_0=1.45$, $\Delta n=1 \times
10^{-4}$, and $\lambda_B =1560$ nm, which are consistent with the
those of Ref.\cite{Longhi2} , the spatial $Z$ and temporal $T$
scales for the example in Fig.1 are $Z \simeq 5$ mm and $ T \simeq
24$ ps, respectively. Hence, in physical units the grating length is
$L \simeq 25$ cm, the phase slips are separated by the distance $aZ
\simeq 1$ cm, and the unit scale of (nonangular) frequency detuning
$E$ from the reference Bragg frequency in Fig.1(c) is $1/ /(2 \pi T)
\simeq 6.6$ GHz.
\begin{figure}[htb]
\centerline{\includegraphics[width=8.2cm]{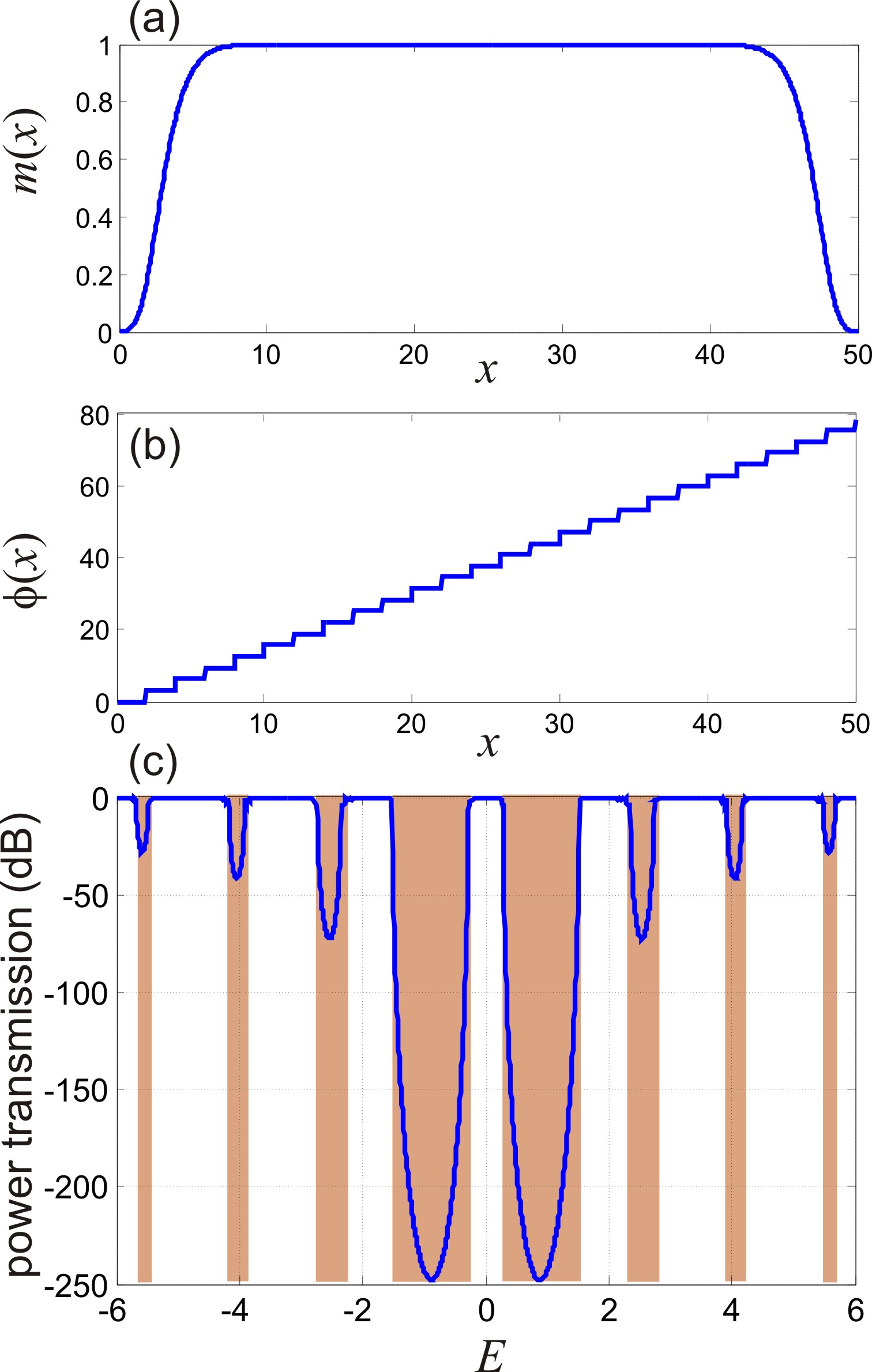}} \caption{ Photonic
analogue of the Dirac-Kronig-Penney model in a superstructure FBG
comprising a periodic sequence of phase slips. (a) and (b):
amplitude and phase profiles of the grating. (c)
Numerically-computed spectral power transmission. Parameter values
are given in the text. The dashed areas in (c) are the stop bands of
the corresponding Dirac-Kronig-Penney infinite lattice.}
\end{figure}
\section{Relativistic Tamm surface states}
The existence of surface Tamm states for the relativistic
Kronig-Penney model has been widely studied in earlier papers by
several authors \cite{D0,D4,SS1,SS2,SS3,SS4,SS5,SS6}. In such
studies, a debate was raised about the proper boundary conditions
that should be imposed to the relativistic wave function at a
$\delta$ barrier (see, for instance, \cite{SS4,BB1,BB2}). As earlier
works \cite{D0,SS1,SS2,SS3}, based on incorrect boundary conditions,
suggested that the relativistic treatment yields a new class of
surface states (the so-called Dirac surface states) which do not
correspond the common Tamm states \cite{Tamm} in the
non-relativistic limit, it was subsequently realized that
application of more physical boundary conditions does not yield any
surface state which violates the Tamm condition in the
non-relativistic limit \cite{SS4}. In our photonic system, the
appropriate boundary conditions in presence of discontinuities of
the phase $\phi(x)$ [and hence a $\delta$-like behavior of the
potential $V(x)$; see Eq.(9)] are readily obtained from Eqs.(3) and
(4), and coincide with those used by  Subramanian and Bhagwat in the
study of relativistic Tamm states \cite{SS4}. The relativistic
extension of the Tamm model is defined by the potential (see, for
instance, \cite{D4,SS4})
\begin{equation}
V(x)= \left \{
\begin{array}{cc}
V_1 & x <0 \\
V_0 \sum_{n=1}^{\infty} \delta(x-na) & x>0
\end{array}
\right.
\end{equation}
Surface states are found as localized solutions to Eq.(8), near the
surface $x=0$, satisfying the appropriate boundary conditions. The
energies of such states, if any, should obviously fall in a gap of
the energy spectrum of the infinitely-extended Dirac-Kronig-Penney
model. As in Ref. \cite{SS4}, we limit to consider the case
$V_1<m_0$, for which surface states occur at the energies $E$ in the
interval $(m_0,m_0+V_1)$ satisfying the equation
\begin{equation}
\kappa {\rm cotg}(\kappa a)=V_1 {\rm cotg}(V_0)-K
\end{equation}
provided that $E-V_1-K {\rm cotg}(V_0)>0$ \cite{SS4}, where $\kappa$
is defined by Eq.(13) and
\begin{equation}
K=\sqrt{m_0^2-(E-V_1)^2}.
\end{equation}
\begin{figure}[htb]
\centerline{\includegraphics[width=8.2cm]{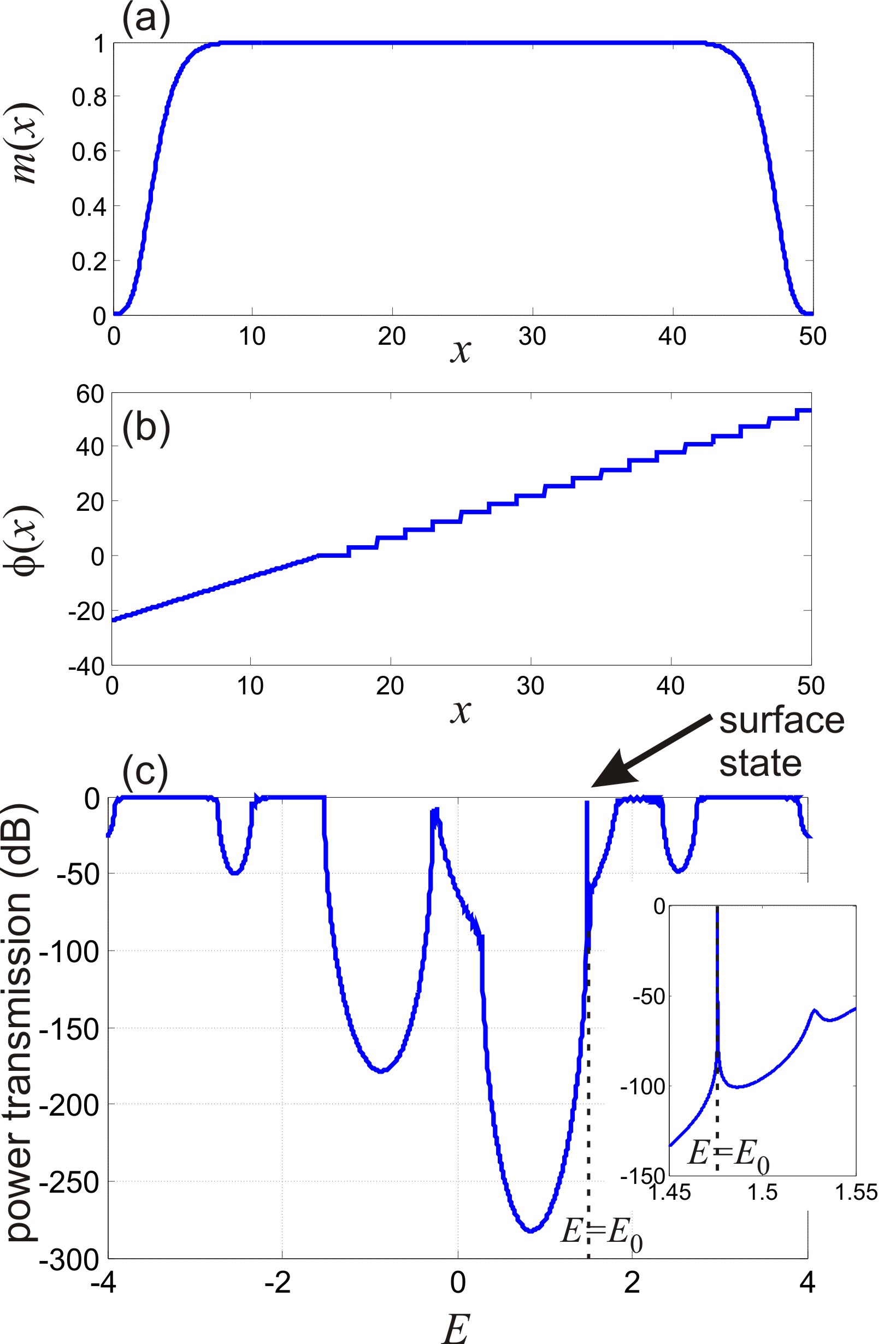}} \caption{ Photonic
analogue of relativistic Tamm surface states in a FBG. (a) and (b):
amplitude and phase profiles of the grating. (c)
Numerically-computed spectral power transmission. The arrow in (c)
is the spectral resonance associated to the surface Tamm state. The
inset shows an enlargement of the spectral transmission spectrum
near the surface-state resonance. Parameter values are given in the
text.}
\end{figure}
In our photonic system, the potential $V(x)$ defined by Eq.(15) and
supporting the surface Tamm states at the $x=0$ boundary corresponds
to a phase profile $\phi(x)$ of the FBG which is composed by a
linearly increasing part with a slope $2V_1$ for $x<0$, and by a
staircase phase profile for $x>0$ [see, for instance, Fig.2(b) to be
commented below], as one can see after integration of Eq.(9),
$\phi(x)=2 \int^x d \xi V(\xi)$. Physically, such a FBG basically
corresponds to two adjacent sections of uniform grating regions but
with different grating periods, with the second grating region (at
$x>0$) comprising a sequence of equally-spaced phase slips, at a
distance $a$, equal to $\Delta \phi=2V_0$. Similarly to the photonic
realization of the Dirac-Kronig-Penney model discussed in the
previous section, a finite grating length is introduced by
apodization of the amplitude profile $m(x)$. In this case, the
surface states actually become resonances of the FBG, i.e. trapped
light states near $x=0$ which are weakly coupled to the external
regions of the fibre because of evanescent tunnelling (see, for
instance, \cite{Longhi10OL}). The existence of surface states can be
thus simply recognized by the appearance of narrow resonance peaks
embedded in a stop band region of the transmission spectrum of the
grating. An example of the photonic analogue of relativistic Tamm
states in a FBG is shown in Fig.2, corresponding to parameter values
$V_0=\pi/2$, $a=2$, $m_0=1$, (same as in Fig.1) and $V_1=0.8$. For
such parameter values, a numerical analysis of the Tamm condition
[Eq.(16)] indicates that there exists one Tamm surface state at the
energy value $E=E_0 \simeq 1.474$, and thus a resonance peak in the
spectral power transmission of the grating is expected at such a
detuning value. Figures 2(a) and 2(b) show the amplitude [Fig.2(a)]
and phase [Fig.2(b)] profiles of the FBG, whereas the corresponding
power transmission (numerically-computed by the transfer matrix
method) is depicted in Fig.2(c). Note that, according to the
theoretical analysis, a strong and narrow resonance peak, at the
detuning $E=E_0$, is clearly observable in the transmission spectrum
[see the inset of Fig.2(c)], which is the signature of the surface
state localized between the two grating regions at around $x=0$. For
parameter values $n_0=1.45$, $\Delta n=1 \times 10^{-4}$, and
$\lambda_B =1560$ nm (as in Fig.1), the spatial $Z$ and temporal $T$
scales are given by $Z \simeq 5$ mm and $ T \simeq 24$ ps,
respectively, and hence in physical units the grating length is $L
\simeq 25$ cm, the phase slips are separated by the distance $aZ
\simeq 1$ cm, and the unit scale of (nonangular) frequency detuning
$E$ from the reference Bragg frequency in Fig.2(c) is $1/ /(2 \pi T)
\simeq 6.6$ GHz.

\section{Conclusions}
In conclusion, in this work a photonic realization of the
Dirac-Kronig-Penney model, describing the band structure of a
periodic potential in the relativistic regime, as well as of
relativistic surface Tamm states, has been proposed. Our photonic
analogue is based on Bragg scattering of light waves in a uniform
FBG with a periodic sequence of phase slips. Band structure and
surface states of the relativistic lattice model can be simply
observed from spectrally-resolved transmission measurements of the
FBG. Design parameters of the grating structures, which are
compatible with the current FBG writing technology \cite{Longhi2},
have been presented.\\

This work was supported by the italian MIUR (PRIN-2008 project
"Analogie ottico-quantistiche in strutture fotoniche a guida
d'onda").

%

\end{document}